# Highly conductive charge transport layers impair charge extraction selectivity in thin-film solar cells

Mathias Nyman[1]*, Christian Ahläng[1], Staffan Dahlström[1], Manasi Pranav[2,§], Johannes Benduhn[2], Syeda Qudsia[3], Jan-Henrik Smått[3], Donato Spoltore[2,+], and Ronald Österbacka[1]

[1]Physics, Faculty of Science and Engineering, Åbo Akademi University, Henriksgatan 2, 20500 Turku, Finland

[2]Dresden Integrated Center for Applied Physics and Photonic Materials (IAPP) and Institute for Applied Physics, Technische Universität Dresden, Nöthnitzer Straße 61, 01187 Dresden, Germany

[3]Laboratory of Molecular Science and Engineering, Faculty of Science and Engineering, Åbo Akademi University, Henriksgatan 2, 20500 Turku, Finland

§ Present address of Manasi Pranav: Institute of Physics and Astronomy, Potsdam University, Karl-Liebknecht-Str. 24-25, 14476 Potsdam-Golm, Germany

+ Present address of Donato Spoltore: Institute for Materials Research (IMO-IMOMEC), Hasselt University, Wetenschapspark 1, 3590 Diepenbeek, Belgium

## Abstract

Charge selective interlayers are crucial in thin-film photovoltaics, such as organic and Perovskite solar cells. Charge transporting layers (doped and undoped) constitute perhaps the most important class of charge selective interlayers; however, it is not well understood how a charge transporting layer should be designed in order to ensure efficient extraction of majority carriers while blocking minority carriers. This work clarifies how well charge-transporting layers with varying majority carrier conductivities block minority carriers. We use the Charge Extraction by a Linearly Increasing Voltage technique to determine the surface recombination velocity of minority carriers in model system devices with varying majority carrier conductivity in the transporting layer. Our results show that transporting layers with high conductivity for majority carriers do not block minority carriers - at least not at operating voltages close to or above the built-in voltage, due to direct bi-molecular recombination across the transporting layer-absorber layer interface. We furthermore discuss and propose design principles to achieve selective charge extraction in thin film solar cells using charge transporting layers.

**Introduction**

Solar cells based on low-mobility semiconductors such as organic polymers and small molecules and hybrid organic-inorganic Perovskites hold great potential for future low-cost energy production. Due to improved understanding of the efficiency limiting factors and better performing materials, the field has seen a rapid increase in power conversion efficiencies in recent years.[1] The improved performance can partly be attributed to reduced charge carrier recombination in the bulk of the active layers observed in many systems (see [2-4] and references therein). As bulk recombination decreases, surface recombination, i.e., the extraction of minority carriers (holes at the cathode and electrons at the anode) becomes an increasingly important loss mechanism.

In order to ensure selective extraction of charge carriers, i.e., holes at the anode and electrons at the cathode, charge selective interlayers are typically introduced. There are plenty of materials available for use as charge selective interlayers, such as doped and undoped organic semiconductors and various metal oxides, to name a few.[5] The thicknesses of the selective layers can vary greatly, from sub-nanometer insulating layers to doped layers 100 nm or thicker. The importance of charge selective interlayers in solar cells can hardly be exaggerated; improving the selectivity of contacts has improved both $V_{OC}$ and FF. [6-10] It has been suggested that highly selective interlayers can be sufficient to provide a diode behavior to the device. In this case, there would be no need for a built-in voltage ($V_{bi}$) [11]. This means that one could make an efficient solar cell even with symmetric work function contacts, allowing for expensive metals such as gold to be replaced by cheaper alternatives.

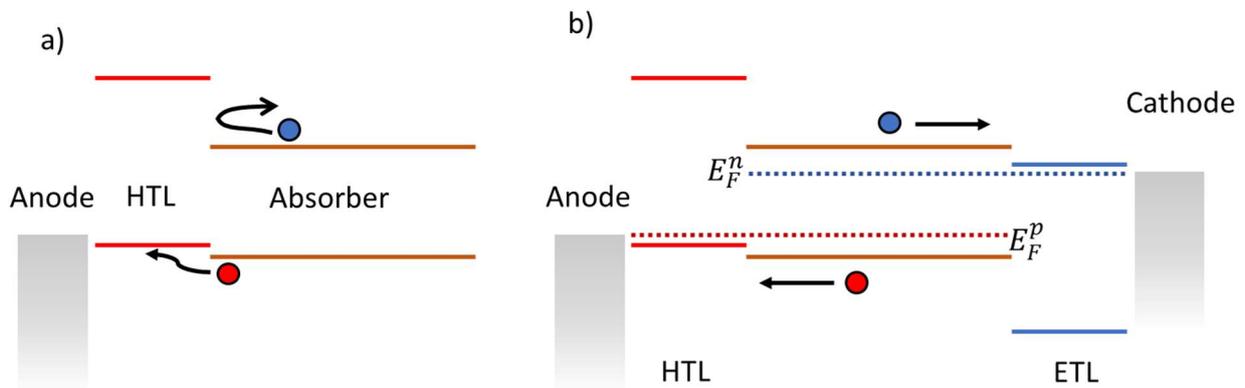

**Figure 1 a)** A hole transporting layer (HTL) between the absorber and the contact blocks the extraction of electrons. **b)** Idealized schematic of the electron and hole quasi-Fermi levels ($E_F^n$ and $E_F^p$) in relation to the contact work functions and energy levels of the HTL and electron transporting layer (ETL).

The purpose of a charge selective interlayer is to efficiently extract one type of charge carrier while blocking the other, thereby suppressing surface recombination (**Figure 1 a)**). Typically this is achieved by choosing a material with a larger bandgap than the light-absorbing material such that the valence bands (or conduction bands) are aligned, thus creating an energy barrier at the conduction bands (valence bands), see **Figure 1 b)**. Figure 1 b) represents an ideal situation under flat-band conditions, trap-free and without photo-generated or equilibrium charges. In this case, it is clear that a photo-generated electron (hole) cannot be extracted at the anode (cathode) due to the high energetic barrier. As such, it is bound to be extracted at the cathode (anode) or recombine in the bulk. However, such an idealized picture is typically not valid in a real device under operation.

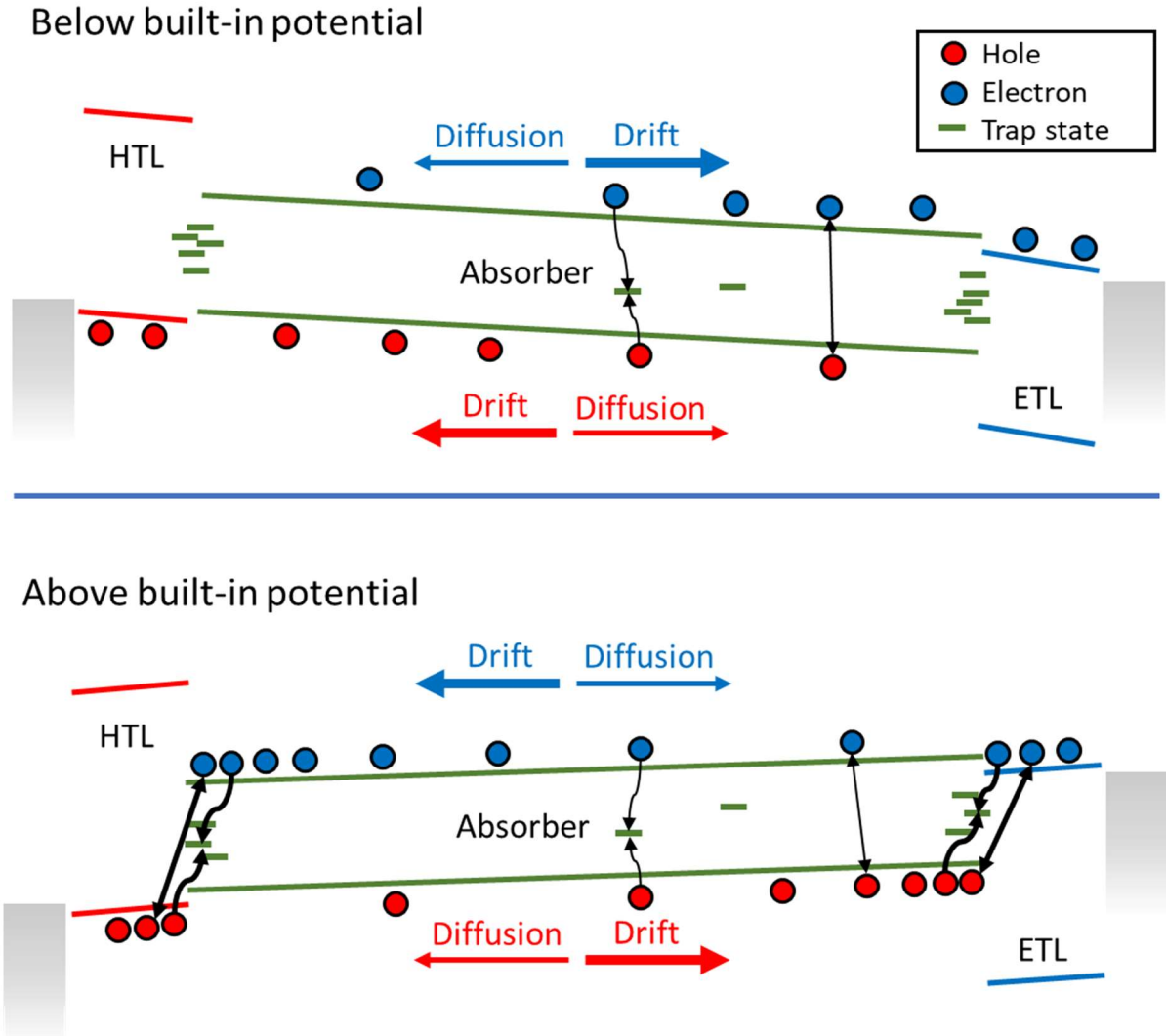

**Figure 2** Energy level schematic of a stack consisting of a hole transporting layer (HTL), absorber and electron transporting layer (ETL), with an applied voltage below (upper panel) and above (lower panel) $V_{bi}$.

A more realistic schematic is shown in **Figure 2**, which includes photo-generated charges both in the absorber and selective layers and charge traps in the absorber and at the absorber/selective layer interface. Since the transport of charge carriers is governed by drift and diffusion, it is elucidating to separately look at the situation when the applied voltage over the device, $V_{appl}$, is below or above the built-in voltage, $V_{bi}$ (since the drift-component goes in different directions). When $V_{appl} \ll V_{bi}$, the internal electric field in the active layer will drive photo-generated electrons towards the cathode and holes towards the anode. In this case, surface recombination will be negligible as the density of minority carriers in the vicinity of the contacts is low. However, when $V_{appl} \gg V_{bi}$, the applied voltage will drive the photo-generated charges in the opposite direction, resulting in significant densities of minority carriers at the contact and concomitant surface recombination. Charges can still be extracted at the correct contact through diffusion (provided that the selective layers block minority carriers). It should be stressed that the actual charge carrier distributions will depend on several factors, such as the charge generation profile, charge carrier recombination rates, and transport properties. Especially at applied voltages close to the built-in potential, it is challenging to determine the actual charge carrier densities.

In most cases, the selective layers have been developed by a trial-and-error approach, and it is not conclusively clarified how a charge selective interlayer should be designed for optimal performance. What is clear is that the efficient extraction of majority carriers is of primary importance. The conductivity, defined as $\sigma = q\mu N$, where $q$ is the electron charge, $\mu$ mobility and $N$ number of charges, for majority carriers (in the selective layers) needs to be high enough such that no, or a minimal, voltage drop occurs across the selective layer. This is achieved by having a high enough mobility or doping concentration or making the selective layer thin enough [7, 8]. However, the processes of minority carrier surface recombination are still not well understood, and consequently, it is not clear how this surface recombination can be suppressed while maintaining a high extraction efficiency for majority carriers. It is generally assumed that trap-assisted recombination contributes to surface recombination; minority carriers either recombine with trapped majority carriers or are themselves trapped to subsequently recombine with free majority carriers [12, 13]. However, it is unclear how the direct recombination (of free minority carriers in the active layer with free majority carriers in the selective layer) contributes to surface recombination. We note that the interface recombination (direct bi-molecular recombination across the interface) for transport layer-active layer interfaces could be significant.

At an interface where electrons are minority charge carriers, the recombination current can be expressed as

$$j_R = -qS_R(n_S - n_0) \qquad (1)$$

where $q$ is the elementary charge, $S_R$ is the surface recombination velocity, $n_S$ the electron concentration at the interface and $n_0$ the equilibrium carrier concentration. In the case of (indirect) trap-assisted recombination, $S_R$ depends on the thermal velocity of electrons, capture cross-section, and density of the interface trap states but is independent of the concentration of both charge carrier types [14]. If the recombination across the interface is direct, then the recombination current depends on the product of both charge carriers,

$$j_R \propto n_S p_S \qquad (2)$$

where $p_S$ is the hole concentration at the interface. Assuming that $j_R \approx -qS_R n_S$, $S_R$ must then be proportional to $p_S$. Hence, by determining $S_R$, it is possible to clarify whether or not direct recombination contributes to surface recombination for a given interface.

In this paper, we clarify how the surface recombination velocity of minority charges depends on the conductivity (of majority carriers) of the charge transporting layer. We use the Charge Extraction by a Linearly Increasing Voltage (CELIV) technique to determine the surface recombination velocity of minority carriers in model devices based on vacuum evaporated small organic molecules. The conductivity in the transporting layer is controlled both by using materials with different mobility (intrinsic semiconductors) and by varying the doping concentration. Our results indicate that the direct bi-molecular recombination between minority carriers in the active layer with majority carriers in the transporting layer can be significant and should not be overlooked. Similar results are also shown for selective layers based on metal oxides. Based on our findings, we present two different design pathways to improve the selective extraction of charges in solar cells based on low-mobility materials.

## Results

Firstly, we clarify how varying the charge carrier mobility in the transporting layer affects the surface recombination velocity of minority carriers. For this purpose, we use the following device structure: **ITO (90 nm) / HTL p-doped with 10 wt% NDP9 (p-dopant no. 9, Novaled GmbH, Germany) (10 nm) / HTL (10 nm) / $C_{60}$ (50 nm) / BPhen (8 nm) / Al (100 nm)**, where ITO is Indium Tin Oxide, Doped HTL is a highly doped hole transporting layer (high enough doping to facilitate Ohmic hole injection), HTL is the same hole transporting layer, but undoped, and BPhen is Bathophenanthroline. As HTLs we use materials with varying energy levels, resulting in different charge transfer state energies $E_{CT}$ (with $C_{60}$ as acceptor), and varying mobilities [15-18] see **Table 1**.

| Material | Mobility [cm$^2$/Vs] | $E_{CT}$ [meV] |
|---|---|---|
| Rubrene | $8 \cdot 10^{-3}$ | 1468 |
| TAPC | $3 \cdot 10^{-4}$ | 1441 |
| Spiro-MeO-TPD | $1 \cdot 10^{-4}$ | 1110 |
| m-MTDATA | $3 \cdot 10^{-5}$ | 996 |

**Table 1** The room temperature mobilities (obtained by time-of-flight technique) and charge transfer state energies ($E_{CT}$, when $C_{60}$ is used as an acceptor) of the HTL-materials used. The chemical formulas of the HTLs are given in the Experimental-section. The mobilities are taken from Refs [16-18], whereas $E_{CT}$ are obtained from sensitive external quantum efficiency spectra as shown in Figure S1 in the Supporting Information.

The surface recombination velocity of minority carriers, in this case electrons, is determined using CELIV.[19] Electrons are injected from the BPhen/Al contact and driven towards the HTL using a DC offset voltage $V_{OFF}$ (larger than the built-in voltage) in forward bias. The electron reservoir is then extracted by a linearly increasing voltage pulse in reverse bias of slope $A = V_{max}/t_{pulse}$, where $V_{max}$ is the maximum applied voltage and $t_{pulse}$ is the length of the pulse. The resulting extraction current density consists of two parts, a time-independent one, $j(0)$, due to the charging of the geometric capacitance, and a time-dependent one $\Delta j(t)$, due to the extraction of the injected charge reservoir. The extracted charge $Q_{extr}$ is given by:

$$Q_{extr} = \int_0^{t_{pulse}}[\Delta j(t) - j(0)]dt \qquad (3)$$

The surface recombination velocity $S_R$ is then given by:

$$S_R = \frac{2\varepsilon\varepsilon_0 kT}{qQ_{extr}^2}J_D \qquad (4)$$

where $\varepsilon\varepsilon_0$ is the dielectric constant, $k$ is the Boltzmann constant, $T$ is temperature, and $J_D$ is the steady-state current before the extraction pulse. Due to the high electron mobility in $C_{60}$, measurements were performed at low temperatures (50 or 100 K) in order to avoid RC effects (the time when the current transient reaches its maximum, $t_{max}$ should be $\gg RC$, where $R$ is the resistance of the outer circuit and $C$ is the capacitance of the device).

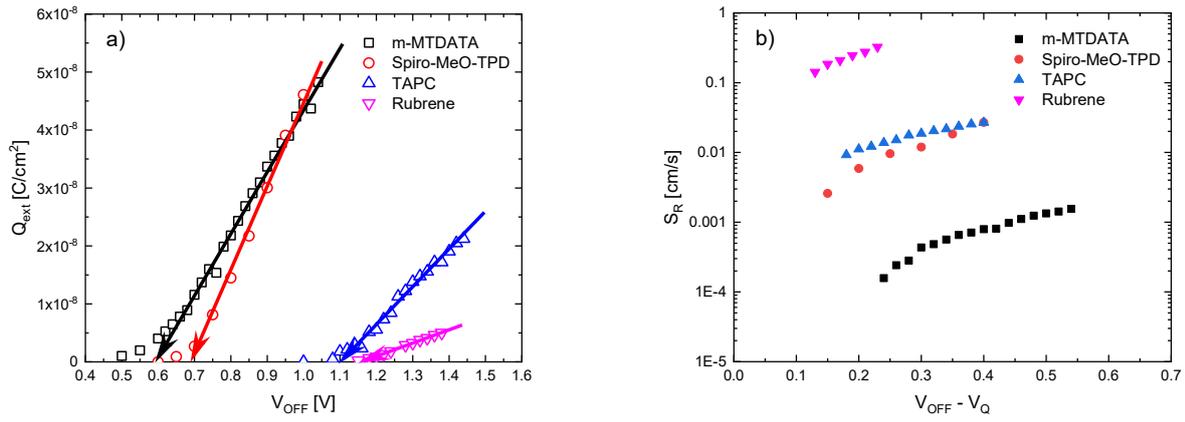

**Figure 3** a) The extracted charges as a function of $V_{OFF}$ in devices with varying HTLs. The arrows are guides to the eye, indicating $V_Q$ (an estimate of $V_{bi}$). b) The corresponding $S_R$ as a function of $V_{OFF}$-$V_Q$.

**Figure 3 a)** shows the extracted charge as a function of $V_{OFF}$ for devices with four different HTLs, the corresponding extraction current transients are given in the **Supporting Information**. It can be seen that at low $V_{OFF}$ there are no injected charge reservoirs since $V_{OFF} < V_{bi}$. When $V_{OFF}$ is large enough, the extracted charge increases linearly with $V_{OFF}$. By then extrapolating to zero extracted charge (indicated by the arrows in Figure 3 a)), one can obtain the onset voltage of charge injection $V_Q$, which roughly corresponds to $V_{bi}$ (assuming an Ohmic cathode). The differences in the obtained $V_Q$ scale with $E_{CT}$ are due to the difference in the quasi-Fermi levels for holes (indirectly the different Highest Occupied Molecular Orbitals of the HTL).

**Figure 3 b)** shows the $S_R$ corresponding to $Q_{ext}$ in Figure 3 a) as a function of $V_{OFF} - V_Q$ (in order to make the comparison between different materials more straightforward). The first thing to note is that the $S_R$ for the Spiro-MeO-TPD device and the TAPC device are almost identical, despite a significant difference in $E_{CT}$. The hole mobility, however, is very similar in Spiro-MeO-TPD and TAPC (see Table 1) [16, 18]. Rubrene, which has similar $E_{CT}$ as TAPC but significantly higher hole mobility, has a roughly one order of magnitude higher $S_R$ [17]. m-MTDATA, on the other hand, has the lowest hole mobility, and indeed the m-MTDATA-device also has the lowest overall $S_R$ [18]. It is clear that for the devices studied in Figure 3, an increase in the conductivity for majority carriers in the HTL by increased hole mobility correlates with an increased $S_R$ for electrons at the HTL-active layer interface.

Another way to increase the conductivity for majority carriers in a transporting layer is by doping. In the following, we clarify the effect of varying doping concentrations in the HTL. For this purpose, we use TAPC with varying wt % of NDP9 as HTL and a $C_{60}$:TAPC low-donor-content (6 wt %) active layer, with the device structure **ITO (90 nm) / HTL (20 nm) / $C_{60}$:TAPC (100 nm) / BPhen (8 nm) / Al (100 nm)**. We have also included devices with a 5 wt % doped TAPC layer and a 5 or 10 nm intrinsic TAPC layer. The results are summarized in **Figure 4**, the corresponding extraction current transients are found in the **Supporting Information**. Even though the measurements were conducted at 50 K, it was not possible to determine the $S_R$ in devices with a higher doping concentration than 2 wt %. We also measured devices with 4 wt %, 5 wt %, 7 wt % and 10 wt % dopant concentration in the HTL. However, $S_R$ was too high to measure since no charge reservoir build-up could be seen. This means that $S_R$ is larger than the effective bulk-limited transport velocity $v_D \approx \frac{\mu kT}{qd}$ where $d$ is the $C_{60}$:TAPC-layer thickness; i.e. electrons recombine at the HTL-active layer interface faster than they can be injected from the cathode and transported to the HTL. From the $t_{max}$ the mobility is ~ 2·10$^{-2}$ cm$^2$/Vs which gives $v_D$ ~ 10 cm/s [20].

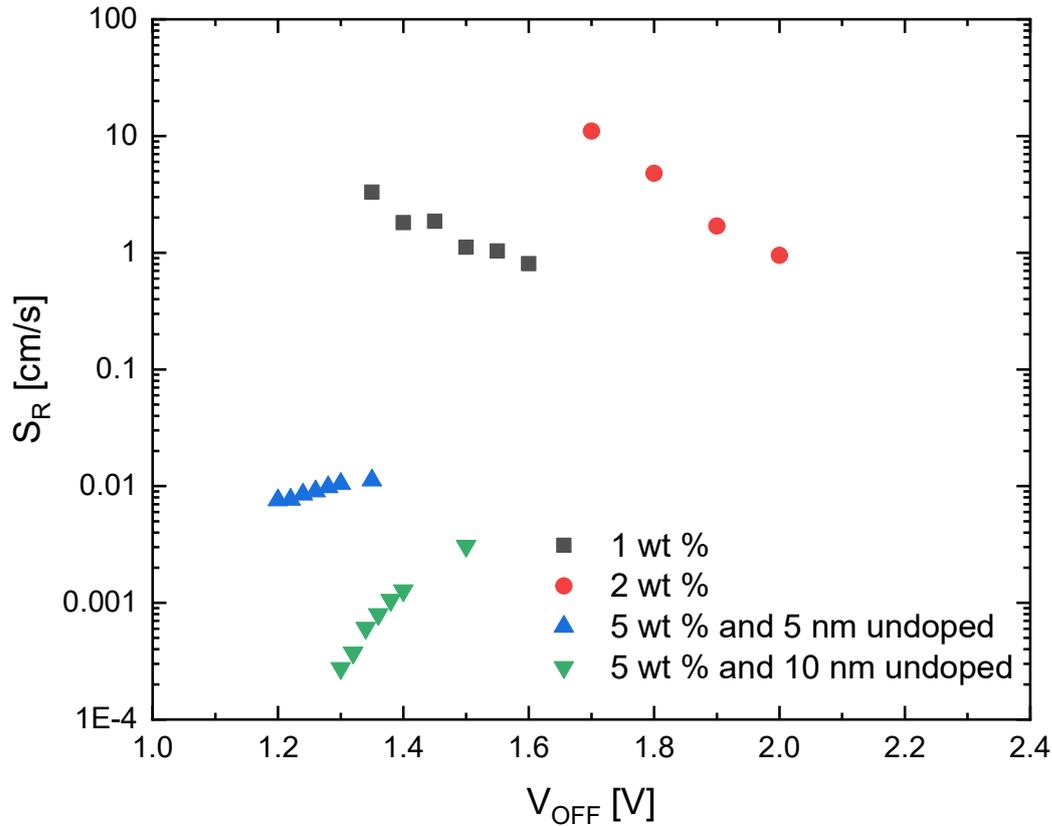

**Figure 4** $S_R$ as a function of $V_{OFF}$ for TAPC-devices with different doping levels in the HTL.

An increase in the doping concentration in the HTL leads to a higher $S_R$ for electrons at the HTL/active layer interface. The insertion of an intrinsic HTL between the doped HTL and the active layer drastically decreases $S_R$, effectively acting as a "passivating layer".

It is noteworthy that $S_R$ is not constant with varying $V_{OFF}$, as expected for purely trap-assisted recombination. In the case of undoped HTLs, $S_R$ increases with increasing $V_{OFF}$ whereas the opposite trend is seen for doped HTLs. Due to direct bi-molecular recombination, $S_R$ depends on the density of holes at the HTL/active layer interface. In the undoped case, the density of holes close to the HTL/$C_{60}$ interface will be highly dependent on $V_{OFF}$. For $V_{OFF} < V_{bi}$, no (or very few) holes are injected from the anode, and the hole density in the HTL is close to zero. For $V_{OFF} > V_{bi}$, holes are injected from the anode and driven towards the HTL/$C_{60}$:TAPC interface where they can recombine with electrons injected from the cathode, the larger the $V_{OFF}$, the larger the hole density in the HTL and the larger $S_R$.

On the other hand, in the doped case, a depletion region forms at the anode when $V_{OFF} > V_{bi}$ with increasing width for increasing $V_{OFF}$. Under steady-state conditions, injected electrons (in $C_{60}$) recombine with doping-induced holes in the HTL. Holes then need to be replenished from the anode. However, injection of holes into the HTL will be limited by the low hole conductivity in the depletion region, effectively leading to a decreasing $S_R$ for increasing $V_{OFF}$ (since the depletion region width increases).

The above results show that an increase in the conductivity for majority carriers in the transporting layer leads to an increased surface recombination velocity of minority carriers due to direct bi-molecular recombination across the interface - at least in devices based on organic semiconductors. Another large class of materials used as selective layers is metal oxides. Most metal oxides are prone to a so-called light-soaking effect, i.e. the work function changes when exposed to UV-light, making them interesting model systems for clarifying surface recombination effects [21]. It has been shown that organic solar cells employing $TiO_2$ as an electron selective layer exhibit a severe s-shape in pristine devices, which disappears with exposure to UV-light. Figure 5 shows the extraction current transients before and after UV-light soaking of a hole-only ITO (90 nm) / $TiO_2$ (30 nm) PTB7 (340nm) / $MoO_3$ (10 nm) / Ag (60 nm) device. It can be seen that the $TiO_2$-layer acts as an efficient hole-blocking layer before light-soaking in UV-light, with $S_R = 2.7 \cdot 10^{-6}$ cm/s in agreement with previous results [19]. However, after light-soaking in UV-light, no charge reservoir can be seen, and therefore the $S_R$ cannot be determined. Again $S_R > v_D$ which means that holes are recombining at the $TiO_2$/PTB7 interface quicker than they can be transported through the PTB7 bulk. Note that $V_{bi}$ most likely increases after light-soaking [21]. However, the device shows no hole-blocking properties in any voltage range.

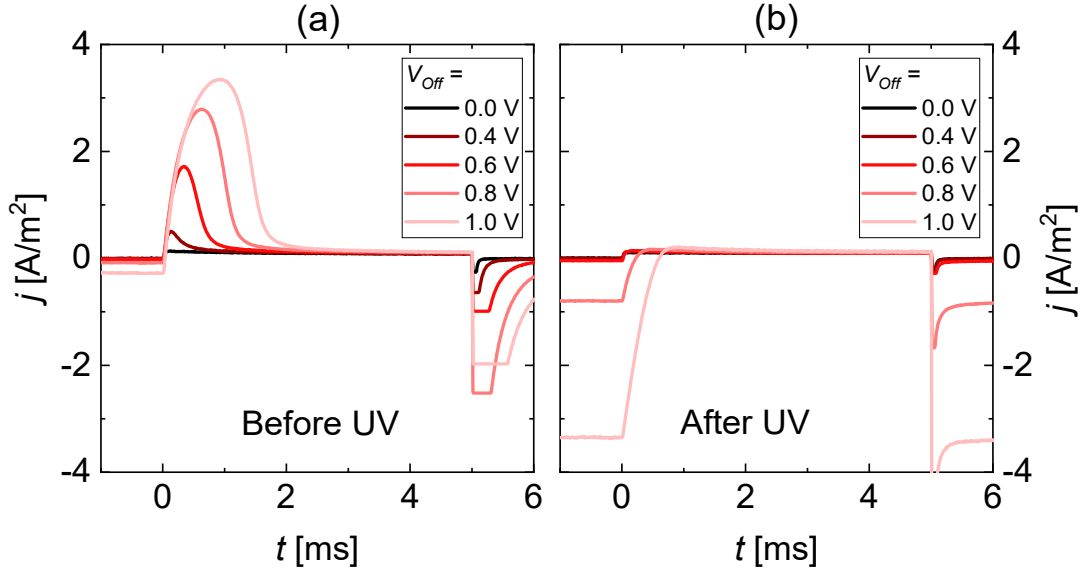

**Figure 5** a) The surface recombination velocity before UV treatment is $S_R = 2.7 \times 10^{-6} \frac{\text{cm}}{\text{s}}$ (average from $V_{\text{Off}} = 0.6$ V to $V_{\text{Off}} = 1.4$ V). b) No reservoir is seen after UV treatment meaning that the TiO$_2$ interlayer acts as a perfect electron collecting contact for the device. The measurement is in this case limited by the effective bulk limited transport velocity $v_D \approx \frac{\mu kT}{qd} \approx 0.08 \frac{\text{cm}}{\text{s}}$, yielding a lower limit to $S_R$ after UV treatment.

The reason why the TiO$_2$-layer effectively does not block holes after exposure to UV-light is most probably due to a drastic increase in the electron density in the TiO$_2$, with a concomitant increase in the direct bi-molecular recombination with injected holes in the active layer. From these results alone, it is not clear what the cause of this increase in electron density is. One possibility is that the photo-catalytic effect causes the TiO$_2$-layer to become highly doped – this would explain both the fact that holes are not blocked at the TiO$_2$/PTB7 interface and the decrease in the effective work function (increase of $V_{\text{bi}}$). Another possibility is that UV-light causes the ITO-work function to decrease, making the ITO/TiO$_2$-contact Ohmic for electrons, thus facilitating electron injection [22].

**Discussion**

The terms selective layers, extraction layers, passivation layers, transporting layers, and blocking layers are used almost interchangeably in the fields of organic and Perovskite photovoltaics. These terms often refer to the same thing; thin interlayers used to facilitate selective charge extraction and suppress surface recombination. However, it is important to remember that these terms are not synonymous. For example, a passivation layer is not necessarily a transporting layer; if the passivation layer is thin enough, charge carriers can tunnel rather than be transported through them.[23, 24] More importantly, as the results presented here demonstrate, a hole transporting (extraction) layer is not the same as an electron blocking layer. As shown, an interlayer with high conductivity for holes effectively does not block electrons due to interface recombination. This has direct consequences for optimizing the selective extraction of charges in real devices.

Surface recombination in organic and Perovskite solar cells is typically assumed to be trap-assisted. In this case, surface recombination can be reduced simply by reducing the density of trap states (reducing the amount of defects etc.). Reducing the number of trap states will always have a positive effect on device performance, regardless of whether the traps are located in the active layer bulk, in the selective layer, or at the interfaces. The critical question is how to reduce the density of trap states. However, we have shown that interface recombination across the absorber layer/transporting layer interface is another potential cause of surface recombination. This means that reducing trap states is a perhaps necessary - but not sufficient requirement to minimize surface recombination. In order to minimize surface recombination, one also needs to minimize interface recombination, which is challenging since a high majority carrier conductivity in the selective layer (necessary for a high FF and $V_{OC}$) can result in significant interface recombination across the selective layer/active layer interface. In the following, we propose two design principles to suppress surface recombination of minority carriers while maintaining an efficient majority carrier extraction.

For any recombination process, the recombination rate $R$ can be expressed as $R = C_r N^\alpha$, where $C_r$ is a recombination constant, $N$ is the density of charge carriers, and $\alpha$ is the reaction order, which depends on the type of recombination in question. For a particular recombination process ($\alpha$ assumed fixed), there are thus two possible ways of reducing $R$; lowering either $C_r$ or $N$.

*Charge carrier distribution – the role of built-in voltage*

The perhaps most straightforward way to reduce interfacial recombination at the selective layer/active layer interface, at least in theory, is to reduce the density of minority carriers at this interface. As discussed earlier, the carrier distributions under operation are governed by drift and diffusion and are a fairly intricate interplay between several parameters such as the generation profile, charge carrier transport, and recombination. This makes it challenging to control the carrier distribution via device design. However, one possible way of achieving a beneficial charge distribution is by increasing $V_{bi}$. The role of $V_{bi}$ in the selective extraction of charges in organic and Perovskite solar cells has been extensively discussed in the literature [7, 11, 24-25]. Our results indicate that hole transporting layers with a high conductivity do not block electrons when $V_{appl} > V_{bi}$. This shows that in devices with charge-transporting layers, the operating voltage should be below $V_{bi}$ to avoid interface recombination, i.e. increasing $V_{bi}$ could make it possible to shift the maximum power point towards higher voltages. This supports the view that a high $V_{bi}$ is beneficial for device performance. However, any positive effect is likely to be highly device-dependent.

*Passivation layers*

Passivation layers of various types have been frequently used in the literature, in particular in Perovskite solar cells. The rationale behind using a passivation layer is that it "passivates" (i.e. de-activates) traps at the selective layer/active layer (or active layer/contact) interface, which results in reduced surface recombination. In the case of trap-assisted recombination, the recombination coefficient is dependent on the trap density (and trap depth), i.e. reducing the trap density reduces the recombination coefficient and thus the recombination rate. However, based on our results, we see another benefit of using a thin interfacial layer between the charge transporting layer and the absorber; suppression of the direct bi-molecular recombination between majority carriers in the transport layer and minority carriers in the absorber. This is demonstrated in Figure 4, where an intrinsic semiconductor layer between the doped HTL and the absorber leads to orders of magnitude reduction in $S_R$. However, the properties of this passivation layer need to be carefully chosen so as not to compromise majority carrier extraction. For example, using an intrinsic transporting layer will not necessarily improve overall device performance (as demonstrated in Figure 4); if the layer is too thick or the mobility too low, majority carrier extraction will be compromised [7, 8]. However, most passivation layers do not actually transport charge. Instead, majority carriers are transferred from the absorber to the transporting layer by tunneling across the passivation layer. This means that a passivation layer always needs to be thin, which might prove to be a challenge for scale-up.

## Conclusions

We have clarified how the surface recombination velocity of minority charges at the charge transporting layer/active layer interface depends on the conductivity (of majority carriers) of the charge transporting layer. Our results show that, due to direct bimolecular recombination across the interface, charge-transporting layers that efficiently transport majority carriers do not block minority carriers. In order to avoid surface recombination in devices with charge-transporting layers, one has to minimize this interfacial recombination, for example, by increasing $V_{bi}$ or employing a thin passivation layer at the transporting layer/active layer interface.

## Acknowledgements

Funding from the Academy of Finland (#32600 and #308307), the Jane and Aatos Erkko foundation (through the ASPIRE project) and the Åbo Akademi researcher mobility program is acknowledged.

J.B. acknowledges the German Federal Ministry of Education and Research (BMBF) for funding through the projects "Pergamon" (16ME0012) and "Flexmonirs" (01DR20008A). DS acknowledges the European Research Council (ERC, grant No. 864625).

## Experimental

**Device fabrication.** The devices were fabricated according to our previous work. The description is reproduced here for completeness.[26] All devices investigated in this work are constructed by a thermal evaporation vacuum system with a base pressure of less than $10^{-7}$ mbar. Before deposition, ITO substrates (Thin Film Devices Inc., USA) are cleaned for 15 min in different ultrasonic baths with NMP solvent, deionized water, and ethanol, followed by $O_2$ plasma for 10 min. The organic materials are purified 1 or 2 times via thermal sublimation. A series of shadow masks and mobile shutters are utilized to control device layout and thickness variation. The effective active area is defined by the geometrical overlap of the bottom and top contact (four different areas were used; 6.44 $mm^2$, 3.22 $mm^2$, 1.61 $mm^2$ and 0.81 $mm^2$). After fabrication, all devices are encapsulated by gluing a transparent glass on top of the device utilizing an epoxy resin (Nagase ChemteX Corp., Japan) cured by UV light. To hinder degradation, a moisture getter (Dynic Ltd., UK) is inserted between the top contact and the glass.

**Materials.** The chemical formulas of the used HTLs are:

| Nr. | Donor (Supplier) | Structure and Chemical Name |
|---|---|---|
| 1 | **Rubrene** (Sensient) | 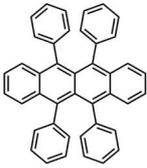 5,6,11,12-tetraphenyl-tetracene |
| 2 | **TAPC** (Sensient) | 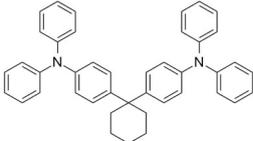 1,1-bis[4-(*N*,*N*-di-*p*-tolylamino)phenyl]cyclohexane |
| 3 | **Spiro-MeO-TPD** (Lumtec) | 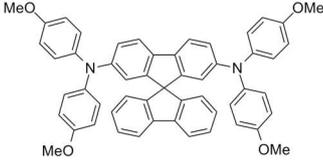 2,7-bis[*N*,*N*-bis(4-methoxy-phenyl)amino]9,9-spiro-bifluorene |
| 4 | **m-MTDATA** (Lumtec) | 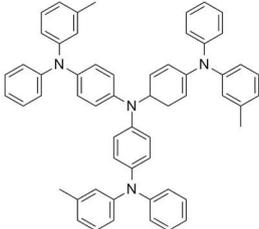 4,4',4"-tris(3-*m*-tolyl-phenylamino)-triphenylamine |

**Sensitive EQE measurements.** The measurements were performed according to previous works, reproduced here for completeness [27]. The light of a quartz halogen lamp (50 W) was chopped at 140 Hz and coupled into a monochromator (Newport Cornerstone 260 1/4 m). The resulting monochromatic light was focused onto the organic solar cell, and its current under short-circuit conditions was fed to a current preamplifier before it was analyzed with a lock-in amplifier (Signal Recovery 7280 DSP). The time constant of the lock-in amplifier was chosen to be 1 s and the amplification of the preamplifier was increased to resolve low photocurrents. The EQE was determined by dividing the photocurrent of the OSC by the flux of incoming photons, which was measured using a calibrated Si and InGaAs photodiode (FDS100-CAL and FGA21-CAL, Thorlabs). According to Ref. [28], the energy of the charge-transfer state $E_{CT}$ was obtained from the low energy tail of the sensitively measured EQE spectra.